\documentclass[10pt,twoside]{hsqcd}
\usepackage{epsf,amsmath}
\usepackage{graphicx}

\begin{document}

\title{RENORMALIZATION PROGRAMME FOR EFFECTIVE THEORIES
\thanks{
This work is supported by INTAS (project 587, 2000), Ministry of
Education of Russia (Programme ``Universities of Russia'') and L.
Meltzers H\o yskolefond (Studentprosjektstipend, 2004).
}}

\author{
\underline{V.VERESHAGIN},
K.SEMENOV-TIAN-SHANSKY, A.VERESHAGIN\\
St. Petersburg State University and University of Bergen\\
E-mail: vvv@av2467.spb.edu
}

\maketitle

\begin{abstract}
\noindent
We summarize our latest developments in perturbative treating the
effective theories of strong interactions. We discuss the principles
of constructing the mathematically correct expressions for the
S-matrix elements at a given loop order and briefly review the
renormalization procedure. This talk shall provide the philosophical
basement as well as serve as an introduction for the material
presented at this conference by A.~Vereshagin and
K.~Semenov-Tian-Shansky
\cite{hsqcd}.
\end{abstract}

\markboth{\large \sl
V.VERESHAGIN et.al.\\
\hspace*{2cm} HSQCD 2004}
{\large \sl \hspace*{1cm}
RENORMALIZATION PROGRAMME FOR EFFECTIVE ...}

\section{Introduction}

In the papers
\cite{AVVV1} -- \cite{AVVV2} we started the systematic study of the
special class of effective theories%
\footnote{
Recall that we use this term in its original (though slightly
modified) meaning suggested by Weinberg
\cite{WeinbEFT}, see also
\cite{WeinMONO}; the precise definitions can be found in
\cite{POMI,AVVV2}.
}
of strong interactions, which we call
{\em localizable}. Roughly speaking, these are theories with the
$S$-matrix which can be obtained in a
{\em perturbative way} in the frame of an
{\em effective field theory}
that contains auxiliary resonance fields along with the fields of
true asymptotic states (those stable with respect to the strong
interaction)%
\footnote{
The localizability property requires separate consideration which we
do not present here due to lack of space. It will be done elsewhere,
the preliminary discussion can be found in
\cite{POMI}.
}.
Our goal is to construct an efficient method for calculating the
amplitudes of physical processes. This means that we need to develop
the systematic scheme of perturbative calculations in the framework of
infinite component effective field theory.

There are few obstacles that usually prevents the effective theory
concept to become a useful computational tool. The main one is the
presence of an infinite number of coupling constants, which requires
introducing an infinite number of renormalization prescriptions. It is
clear that we hardly get some predictive power for our theory unless
find some
{\em regularity} in the system of those prescriptions. Further, if one
admits an unlimited number of the resonance fields in a theory (as we
do), then even the amplitude of a given loop order (say, tree level)
acquires contributions from the infinite number of graphs. The problem
of convergence of the latter
{\em functional} series is the problem of correct definition of the
loop expansion.

Our approach suggests a solution. First, we systematize the set of
renormalization prescriptions. In
\cite{AVVV2}
we demonstrated that only the combinations of coupling constants that
we call as
{\em resultant} parameters require fixing to obtain renormalized
$S$-matrix at any given loop order. Next, we show that even these
renormalization prescriptions cannot be taken arbitrary --- it is the
convergence requirements giving a hand. Here is the basic idea. Any
requirement of convergence of functional series always can be thought
as a
{\em restriction for the parameters}
appearing in those series. In other words,
{\em the couplings, and so the renormalization prescriptions, are
unavoidably restricted for each step of perturbation expansion to make
sense}. It is this circumstance that eventually gives rise to the
system of bootstrap equations for physical parameters (see
\cite{AVVV1,POMI})
and allows to make numerical predictions.

Briefly speaking, we develop Dyson's perturbation scheme for the
infinite component effective theories taking seriously the problems of
mathematical correctness and self-consistency. In this talk we try to
make a short overview of our strategy and explain the main postulates.

\section{Resultant (minimal) parameters and renormalization
prescriptions}

The first (and, probably, the main) step toward a classification of
the renormalization prescriptions is transition to the
{\em minimal parametrization}
discussed in
\cite{AVVV2}.
This allows one to rewrite every given graph in terms of the minimal
propagators and minimal effective vertices. The numerator of the
minimal propagator is just a covariant spin sum considered as a
function of four independent components of momentum. At tree level the
minimal vertices are just the Hamiltonian on-shell vertices with the
wave functions crossed out.

To be precise, one shall consider an
$S$-matrix element of the formal infinite sum of all the Hamiltonian
items constructed from a given set of, say,
$k$ (normally ordered) field operators. The members of this sum differ
from each other by the Hamiltonian coupling constants, by number of
differentiation operators and/or, possibly, by their matrix structure.
The matrix element under consideration should be calculated on the mass
shell, presented in a Lorentz-covariant form and considered as a
function of
$4(k-1)$ independent components of the particle momenta. The wave
functions should be crossed out. The resulting structure we call as the
$k$-leg
{\em minimal effective vertex of the 0-the level}.
Every minimal effective vertex presents a finite sum of scalar
formfactors dotted by the corresponding tensor structures, each
formfactor being a formal power series, or even just a polynomial in
relevant scalar kinematical variables. The
{\em l-th level} effective vertex only differs from that described
above by the presence of bubbles%
\footnote{Some of them may have multi-loop inner structure.}
($l$ loops in total) attached to the same point as the external legs.
The numerical coefficients (eventually supplied with the index
$l$) in the formal power series that describe the corresponding
formfactors are called as the
{\em $l$-th level minimal parameters}. The above-mentioned resultant
parameters are certain sums of the minimal ones.

The resultant parameters are all independent, as far as one considers
as independent all the Hamiltonian couplings. Besides, as shown in
\cite{AVVV2}, every
{\em amplitude graph}%
\footnote{
That computed on the mass shell of all external particles and dotted
by the relevant wave functions.
}
depends only on minimal parameters. In turn, the full sum of
$L$-th loop order graphs describing certain scattering process can be
expressed solely in terms of the resultant parameters with level index
$l \leq L$.

At least a few words should be said here about the renormalization
prescriptions. In
\cite{AVVV2} it is shown that, if the renormalization point is taken on
shell, the resultant parameters are the only quantities that require
formulating renormalization prescriptions. Actually, our use of the
minimal parametrization implies that we rely upon the scheme of
{\em renormalized perturbation theory}. In this scheme one starts from
the physical action written in terms of
{\em physical} parameters and adjusts the counter terms in such a way
that the numerical values of those parameters remain unchanged after
renormalization. It is this fact that allows us to obtain the bootstrap
restrictions for the physical (in principle
{\em measurable}) parameters later on.

\section{Polynomial boundedness and summability}

One of the most important requirements which we make use of when
constructing the meaningful items of Dyson's perturbation series is
that of polynomial boundedness. Namely, the full sum of
$S$-matrix graphs with a given set of external lines and fixed number
$L$ of loops must be polynomially bounded in every pair energy
at fixed values of the other kinematical variables%
\footnote{
This is not precise enough: in the case of several variables one has
first to fix the concrete choice of them. We shall not discuss it in
detail here.
}.
There are two basic reasons for imposing this limitation. First, from
the general postulates of quantum field theory
(see, e.g.,
\cite{axioms})
it follows that the full (non-perturbative) amplitude must be a
polynomially bounded function of its variables. Second, from
experiment it follows that this is quite a reasonable requirement.

Since we never fit data with non-perturbative expressions for the
amplitude, it is natural to impose the polynomial boundedness
requirement on a sum of terms up to any fixed loop order and, hence,
on the sum of terms of each order. Similar argumentation also works
with respect to the bounding polynomial degrees. To avoid unnecessary
mutual contractions between different terms of the loop series, it
makes sense to attract the following
{\em asymptotic uniformity} requirement.
{\em The degrees of the bounding polynomials which specify the
asymptotics of the amplitude of a given loop order must be equal to
those specifying the asymptotics of the full (non-perturbative)
amplitude of the process under consideration.}
Surely, this latter degree may depend on the type of the process as
well as on the values of the variables kept fixed.

The condition of asymptotic uniformity (or, simply, uniformity) is
concerned with the asymptotic behavior of the items corresponding to
different loop orders. It tells us not too much about the rules needed
to convert the disordered sum of graphs with the same number of loops
(and, of course, describing the same process) into the well-defined
({\em summable}) functional series. To solve the latter problem we
rely upon another general principle which we call as
{\em summability requirement}%
\footnote{
By analogy with the maximal analyticity principle widely used in the
analytic theory of S-matrix (see, e.g.
\cite{Chew}) sometimes we call it as
{\em analyticity principle}.
}.
It is formulated as follows.
{\em In every sufficiently small domain of the complex space of
kinematical variables there must exist an appropriate order of
summation of the formal sum of contributions coming from the graphs
with a given number of loops, such that the reorganized series
converges. Altogether, these series must define a unique analytic
function with only those singularities which are presented in
contributions of individual graphs.}

At first glance, the summability (analyticity) requirement may seem
somewhat artificial. This is not true. There are certain mathematical
and field-theoretical reasons for taking it as the guiding principle
that provides a possibility to manage infinite formal sums of graphs
in a way allowing to avoid inconsistencies. It is, actually,
{\em both}
the summability and uniformity principles that allow us to use the
Cauchy formula and obtain well defined expression for the amplitude
of a given loop order
\cite{hsqcd,AVVV1,POMI,AVVV2}.

We would like to stress that the requirements of uniformity and
summability are nothing but independent subsidiary conditions fixing
the type of perturbation scheme which we only work with. Surely, there
is no guarantee that on this way one can construct the most general
expressions for the S-matrix elements in the case of effective theory.
Nevertheless, there is a hope to construct at least a meaningful ones
presented by the Dyson's type perturbation series only containing the
well-defined items.

\section{Amplitude calculation and bootstrap}

The explicit calculations are better illustrated by the concrete
examples
\cite{hsqcd}, here we just briefly sketch the strategy.

One needs to consider any process and classify all the resultant
parameters contributing to it at a given loop order%
\footnote{
So far we performed explicit calculations only for the case of tree
level processes. Although the procedure generalizes for any loop
level, the calculations becomes cumbersome in the latter case. We
shall demonstrate a concrete example of 1 loop calculation in the
forthcoming publications.
}.
Then, using the stated above principles of uniformity and summability,
one applies the well known Cauchy formula to the amplitude of the
given order at various regions in the space of kinematical variables.

Several things happen during this process. First, the summability
principle helps to identify the parameters of the amplitude
singularities as a combination of resultant parameters which are
already fixed by renormalization conditions. With this in hands, and
with the degree of bounding polynomial given by uniformity principle%
\footnote{At this step we attract the known information on the Regge
intercepts.},
the Cauchy formula allows one to write down a
{\em well defined expression for the amplitude in a given domain
(layer) of the space of kinematical variables.}

Next, one obtains
{\em different expressions for the same amplitude in different layers},
different couplings and masses contributing to each of them. The layers
intersect, so one should equate the expressions in the common domains
of validity to ensure self-consistency (usually, it appears to be
equivalent to requirement of crossing symmetry of the given loop
order amplitude).

The latter step gives an infinite number of numerical relations for
the resultant parameters and thus, as explained above, for the
renormalization prescriptions or, the same, for the
physical parameters of a theory. We call these relations as the
{\em bootstrap equations} of a given loop order. It should be noted,
that, although the bootstrap for each loop order amplitude gives
restrictions for
{\em physical} parameters (the latter are independent of the loop
order), the structure of bootstrap equations themselves varies from
order to order. The search for the
{\em solution} of all these equations is still beyond our scope%
\footnote{
See, however, the discussion in
\cite{POMI}.
},
what we can only do so far is to test them with experimental data.
The results are presented in two other talks, see
\cite{hsqcd}.

\section*{Acknowledgments}
We are grateful to V.~Cheianov, H.~Nielsen, S.~Paston, J.~Schechter,
A.~Vasiliev and M.~Vyazovski for stimulating discussions.


\end{document}